\documentclass[onecolumn,superscriptaddress,floatfix,preprint]{revtex4}
\usepackage{graphics,amssymb,amsmath,epsfig,color,braket}
\usepackage{graphicx}
\usepackage[colorlinks, allcolors=blue]{hyperref}
\def\iu{\mathrm{i}}
\def\du{\mathrm{d}}
\begin{document}

\title{Chiral terahertz wave emission from the Weyl semimetal TaAs}
\author{Y. Gao}
\affiliation{State Key Laboratory of Electronic Thin Films and Integrated Devices, University of Electronic Science and Technology of China, Chengdu 610054, China}
\author{Y. Qin}
\affiliation{State Key Laboratory of Electronic Thin Films and Integrated Devices, University of Electronic Science and Technology of China, Chengdu 610054, China}
\author{S. Kaushik}
\affiliation{Department of Physics and Astronomy, Stony Brook University, Stony Brook, New York 11794, USA}
\author{E. J. Philip}
\affiliation{Department of Physics and Astronomy, Stony Brook University, Stony Brook, New York 11794, USA}
\author{Y. P. Liu}
\affiliation{State Key Laboratory of Electronic Thin Films and Integrated Devices, University of Electronic Science and Technology of China, Chengdu 610054, China}
\author{Y. L. Su}
\affiliation{State Key Laboratory of Electronic Thin Films and Integrated Devices, University of Electronic Science and Technology of China, Chengdu 610054, China}
\affiliation{Institute of Electronic and Information Engineering, University of Electronic Science and Technology of China, Dongguan 523808, China}
\author{X. Chen}
\affiliation{Department of Physics and Astronomy, Stony Brook University, Stony Brook, New York 11794, USA}
\author{Z. Li}
\affiliation{State Key Laboratory for Artificial Microstructure and Mesoscopic Physics, Beijing Key Laboratory of Quantum Devices, Peking University, Beijing 100871, China}
\affiliation{Beijing National Laboratory for Condensed Matter Physics, Institute of Physics, Chinese Academy of Sciences, Beijing 100190, China}
\author{H. Weng}
\affiliation{Beijing National Laboratory for Condensed Matter Physics, Institute of Physics, Chinese Academy of Sciences, Beijing 100190, China}
\affiliation{Songshan Lake Materials Laboratory, Guangdong 523808, China}
\author{D. E. Kharzeev}
\affiliation{Department of Physics and Astronomy, Stony Brook University, Stony Brook, New York 11794, USA}
\affiliation{Department of Physics, Brookhaven National Laboratory, Upton, New York 11973-5000, USA}
\affiliation{RIKEN-BNL Research Center, Brookhaven National Laboratory, Upton, New York 11973-5000, USA}
\author{M. K. Liu}
\email{mengkun.liu@stonybrook.edu}
\affiliation{Department of Physics and Astronomy, Stony Brook University, Stony Brook, New York 11794, USA}
\author{J. Qi}
\email{jbqi@uestc.edu.cn} 
\affiliation{State Key Laboratory of Electronic Thin Films and Integrated Devices, University of Electronic Science and Technology of China, Chengdu 610054, China}

%\date{\today}
\begin{abstract}
	As a fascinating topological phase of matter, Weyl semimetals host chiral fermions with distinct chiralities and spin textures. Optical excitations involving those chiral fermions can induce exotic carrier responses, and in turn lead to novel optical phenomena. Here, we discover strong coherent chiral terahertz emission from the Weyl semimetal TaAs and demonstrate unprecedented manipulation over its polarization on a femtosecond timescale. Such polarization control is achieved via the colossal ultrafast photocurrents in TaAs arising from the circular or linear photogalvanic effect. We unravel that the chiral ultrafast photocurrents are attributed to the large band velocity changes when the Weyl fermions are excited from the Weyl bands to the high-lying bands. The photocurrent generation is maximized at near-IR frequency range close to 1.5 eV. Our findings provide an entirely new design concept for creating chiral photon sources using quantum materials and open up new opportunities for developing ultrafast opto-electronics using Weyl physics.	
	\end{abstract}
\maketitle

\section{Introduction}
The generation and control of photoinduced charge current and the resultant electromagnetic wave emission are of crucial importance for coherent operation in opto-electronic quantum devices \cite{Lodahl_Nat_2017}. The merit is two-fold. First, the photocurrents induced by optical transitions obeying the selection rules and/or chirality of the materials naturally permit ultrafast manipulation. This is especially true when the non-thermal excitation of both charge and spin degrees of freedom can be utilized \cite{Kampfrath_NatPhot_2011,Cireasa_NatPhys_2015}. Second, the emission of electromagnetic wave induced by ultrafast currents is essential in the terahertz (THz) frequency regime, where control of the ellipticity and chirality over a broad spectral range is known to be notoriously difficult \cite{Amer_APL_2005,Lu_PRL_2012,WangW_PRL_2015,Zhang_NP_2018,Brixner_PRL_2004,Sato_NP_2013}. Specifically, previous polarization control schemes mainly rely on the sophisticated pulse shaping or two-pulse manipulation of the incident light, where the associated THz emitter itself permits no intrinsic optical chiralites. For example, conventional nonlinear crystals such as ZnTe and LiNbO$_3$ only allow linearly polarized THz emission, irrespective of the polarization of the incident light. On the other hand, a promising way to achieve the polarization control in a wide spectral regime is to exploit the novel topology of band structures where electrons demonstrate unique spin-momentum locking or chiral properties \cite{Hasan_RMP_2010,Qi_RMP_2011,Weng_APhys_2015}. The novel light-matter interactions in quantum materials with topological structures may hold the key to many technical innovations, including ultrafast quantum communication, coherent information processing and controllable data storage \cite{Zutic_RMP_2004,Kirilyuk_RMP_2010}.    

Weyl semimetals (WSMs), with their unique topological electronic structure, retain chiral electrons near the Weyl nodes \cite{Wan_PRB_2011,Burkov_PRB_2011,Vafek_ARCMP_2014,Weng_PRX_2015,Xu_Science_2015,Lv_PRX_2015,Sun_PRL_2016} and, hence, are highly promising for the above-mentioned applications. For example, previous work has proposed that the emergence of various fascinating electronic responses to light is intimately associated with the Berry phase of the topological bands, e.g., spin-polarized photovoltaic currents \cite{Chan_PRB_2017,Zhang_PRB_2018,Ishizuka_PRL_2016}, photoinduced anomalous Hall effect \cite{Chan_PRL_2016}, and quantized photocurrents \cite{Juan_NC_2017}. Therefore the investigation of photocurrents in WSMs raises enormous interest both theoretically and experimentally \cite{Chan_PRB_2017,Zhang_PRB_2018,Ishizuka_PRL_2016,Chan_PRL_2016,Juan_NC_2017,Ma_NPhys_2017,Sun_CPL_2017,Osterhoudt_arXiv_2018}. For mid-infrared light, Refs.~\cite{Chan_PRB_2017,Zhang_PRB_2018,Ma_NPhys_2017} show the existence of a dominant helicity-dependent DC photocurrent due to the circular photogalvanic effect (CPGE) in the WSM TaAs. In contrast, with linearly polarized light, only a giant linear photogalvanic effect (LPGE) (or shift current) was observed in Ref.~\cite{Osterhoudt_arXiv_2018}. For near-infrared light, one work reported the photocurrent measurements \cite{Sun_CPL_2017}, which suggest the existence of the CPGE in TaAs. However, up to now there is very little information about the ultrafast photocurrent in WSMs, except that its time-averaged direction can be derived by considering the crystal symmetry \cite{Ma_NPhys_2017,Sirica_arXiv_2018}. Particularly, properties of the ultrafast photocurrent induced THz emission and its associated Weyl electron dynamics are still unclear.     

\begin{figure*}
	\includegraphics[width=16cm]{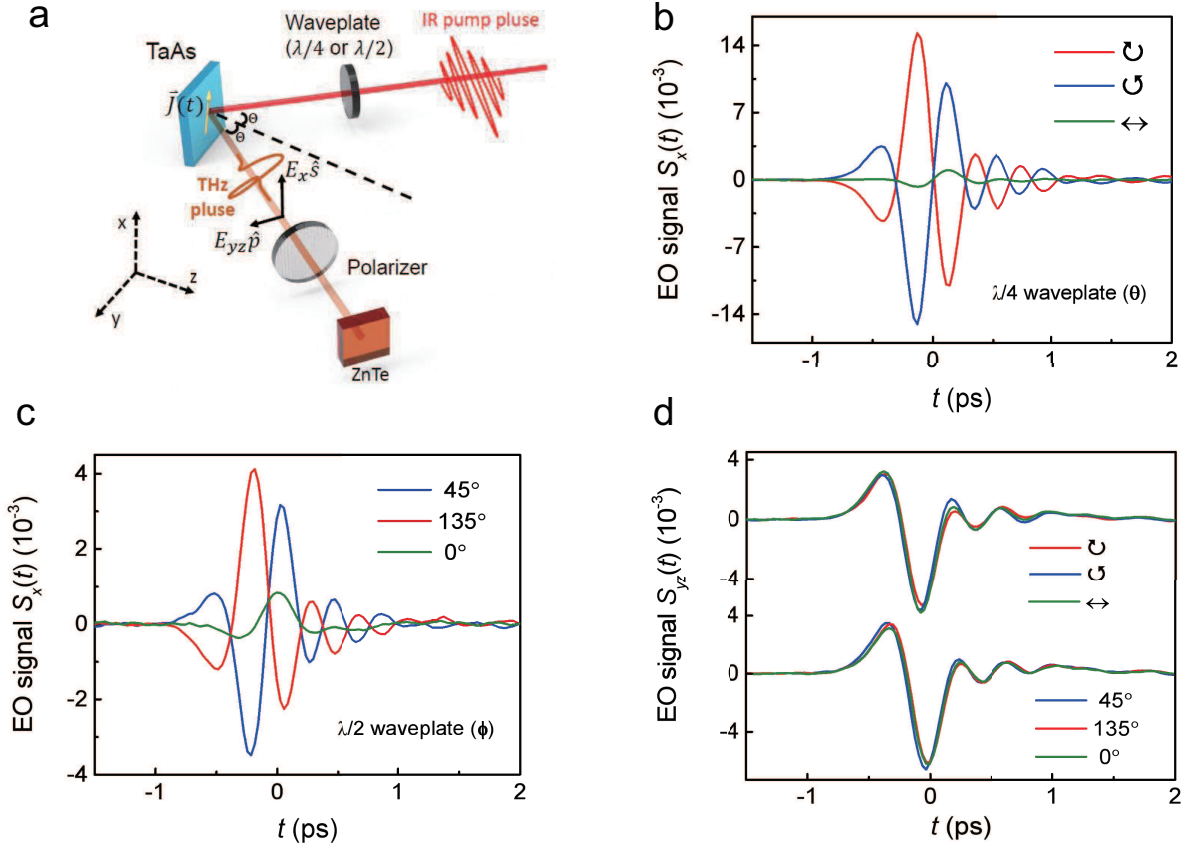}
	\caption{\label{fig:mainTHzresults} \textbf{a}. Schematic of the
		THz emission spectroscopy. Excitation of a fs laser pulse with an incident angle $\Theta$ onto a TaAs single crystal initiates a photocurrent burst and, consequently, emission of a THz pulse $\vec{E}(t)$[$=E_x(t)\hat{s}+E_{yz}(t)\hat{p}$]. Measurement of the components $E_x(t)$ and $E_{yz}(t)$ by the EO sampling provides access to the sheet current density $\vec{J}(t)$ flowing inside the sample. \textbf{b-d}. Typical THz EO signal components $S_x(t)$ and $S_{yz}(t)$ along the $\hat{s}$ and $\hat{p}$ directions were measured at various settings for pump polarization via rotating the $\lambda/4$ or $\lambda/2$ waveplate, characterized by the angle $\theta$ or $\phi$. Here, $\leftrightarrow$ ($\theta$=0$^\circ$), $\circlearrowright$ ($\theta$=45$^\circ$), and $\circlearrowleft$ ($\theta$=135$^\circ$) represent the $p$, right-handed, and left-handed circularly polarized light, respectively. The angle $\phi$ stands for the linear polarization state with respect to the $p$ polarized light ($\phi$=0$^\circ$).}
	%\vspace*{-0.4cm}
\end{figure*}

In this paper, we reveal a new route to realize the generation and control of broadband elliptically polarized THz waves in the Weyl semimetal TaAs. We quantitatively elucidate the colossal ultrafast photocurrents in TaAs for the first time. We find that the polarization of the elliptically polarized THz wave can be easily manipulated on a fs timescale, which is unprecedented. Such control arises from the colossal chiral ultrafast photocurrents whose direction and magnitude can be manipulated in an ultrafast manner using the circularly and/or linearly polarized fs optical pulses. The excitation pulse can have a broad spectral range from visible to mid-infrared light, and generate maximum photocurrent around 1.5 eV. We unravel that the Weyl fermions play the key role in generating the giant chiral ultrafast photocurrents. 

A schematic of the experiment is shown in Fig.~\ref{fig:mainTHzresults}\textbf{a}. Femtosecond laser pulses are used to induce ultrafast photocurrents. According to the Maxwell equations, a change in the current density $\vec{j}(z,t)$ on the picosecond (ps) timescale will result in electromagnetic radiation in the THz spectral range (1 THz = 1 ps$^{-1}$) \cite{Kampfrath_NNano_2013}. The transient electric field $\vec{E}(t)$ is generated with a polarization parallel to the direction of the current. Therefore, one can use the time-domain spectra $\vec{E}(t)$ of the THz radiation as a probe for the ultrafast sheet current density given by $\vec{J}(t)=\int dz\vec{j}(z,t)$. The orthogonal components $J_x$ and $J_{yz}$ via the generalized Ohm's law determine the THz near-field $\vec{E}(t)$ on top of the sample surface, i.e., the $s$-polarized $E_x$ along $\hat{x}$ and the $p$-polarized $E_{yz}$ in the $yz$ plane. Experimentally, the THz far-field electro-optic (EO) signal $\vec{S}(t)$ was collected, and the THz near-field $\vec{E}(t)$ was derived via inversion procedures based on a linear relationship between these two quantities \cite{Kampfrath_NNano_2013} (see Appendix).  Therefore, $\vec{S}(t)$ is a qualitative indicator of the ultrafast photocurrent, whose genuine properties shall be quantitatively obtained by analyzing $\vec{J}(t)$.    

\section{Results and discussions}
\begin{figure}
	\includegraphics[width=16cm]{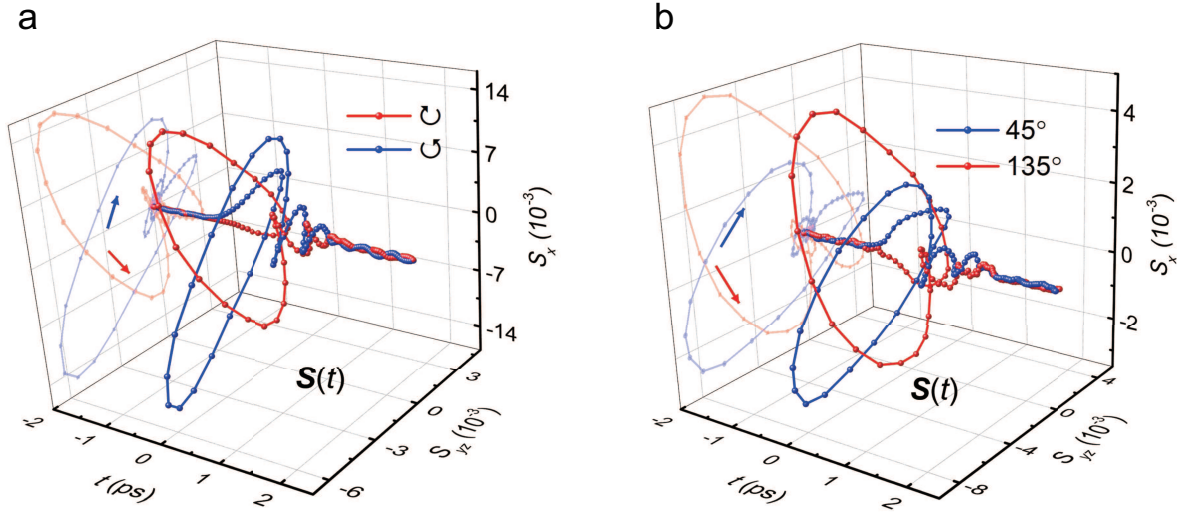}
	\caption{\label{fig:3D} \textbf{a} and \textbf{b} are the far-field EO signals $\vec{S}(t)$ $[=S_x(t)\hat{s}+S_{yz}(t)\hat{p}]$ for circularly and linearly polarized pump light, respectively. The coloured arrows indicate different optical chiralities: left-handed (blue) and right-handed (red).}
	%\vspace*{-0.4cm}
\end{figure}

\textbf{THz emission from TaAs.} In the present experiments, unless noted in the text, we mainly focus on the results obtained for the TaAs(112) single crystal with an incident angle $\Theta\simeq3^\circ$ using the excitation light with a wavelength of 800 nm. $\hat{x}$ is along the [$\overline{1}$10] direction. The results for (011) and (001) faces, another angle of incidence ($\Theta\simeq45^\circ$), and other excitation wavelengths are reported in the Supplemental Material.  Figs.~\ref{fig:mainTHzresults}\textbf{b-d} show strong time-domain THz far-field EO signals $\vec{S}(t)$ detected from the sample. Within our pump power range, the THz peak field strength can reach up to $\sim$1 kV/cm and the dynamic range of $\vec{S}(t)$ can be $\sim$60 dB (see Supplemental Material). Clearly, both the magnitude and temporal shape of the THz waveform $S_x(t)$ depend strongly on the light polarization. The key observation is that signals $S_x(t)$ taken with right- ($\circlearrowright$) and left-handed ($\circlearrowleft$) circularly polarized light are completely out of phase (Fig.~\ref{fig:mainTHzresults}\textbf{b}). A similar observation was found for the 45$^\circ$ and 135$^\circ$ linearly polarized light (Fig.~\ref{fig:mainTHzresults}\textbf{c}). In terms of the peak values, $S_x$ induced by the linearly polarized light is approximately three times smaller than that due to excitation by circularly polarized light. By contrast, $S_{yz}(t)$ is almost polarization-independent and differs substantially from $S_x(t)$ (see Fig.~\ref{fig:mainTHzresults}\textbf{d}). Such distinct $S_x$ and $S_{yz}$ components lead to a elliptically polarized transient THz field $\vec{S}(t)$ (or $\vec{E}(t)$), which exhibits opposite chirality for different circularly or linearly polarized pump light (see Figs.~\ref{fig:3D}\textbf{a} and \textbf{b}), as we will discuss in detail below.

\begin{figure}
	\includegraphics[width=16cm]{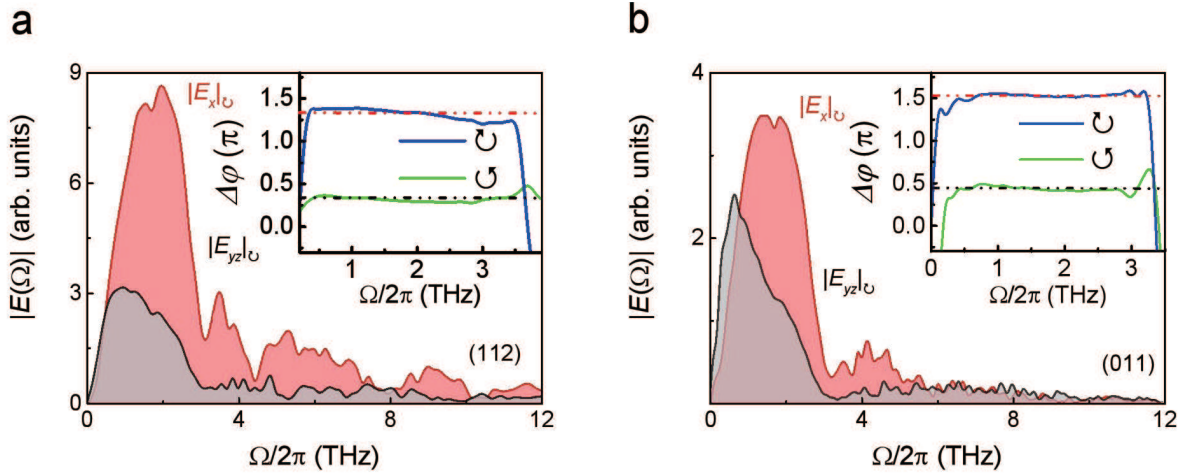}
	\caption{\label{fig:Fourier} \textbf{a} and \textbf{b} are Fourier transform spectra for the THz near-field $\vec{E}(t)$ from (112) and (011) faces for circularly polarized pump light, respectively. The insets show the phase difference ($\Delta\varphi$) between $E_x(\Omega)$ and $E_{yz}(\Omega)$ for each circularly polarized pump light. The dashed lines represent the average value of $\Delta\varphi$.}
	%\vspace*{-0.4cm}
\end{figure}

In the frequency domain of the THz near-field $\vec{E}(t)$ (see Fig.~\ref{fig:Fourier}\textbf{a}-\textbf{b}), the dominant spectra for both $E_x$ and $E_{yz}$ sit below $\sim$3 THz. At approximately 1.7 and 3.1 THz ($\sim$57 and 104 cm$^{-1}$), there exist two obvious dips. The former might be due to the infrared active phonon mode in TaAs. The latter can be attributed to the absorption of the Raman active $E(1)$ mode \cite{Liu_PRB_2015}. The high-frequency tails extend almost to 12 THz, consistent with a time resolution of $\sim$80 fs. Strikingly, for 0.2$\lesssim\Omega\lesssim$3 THz, we discovered that the phase difference, $\Delta\varphi$, between $E_x$ and $E_{yz}$ is nearly constant. This phase difference is independent of the incident angle $\Theta$ for a given pump polarization. However, its value differs between different faces (insets of Figs.~\ref{fig:Fourier}\textbf{a} and \textbf{b}), i.e. for (112), $\Delta\varphi_\circlearrowleft\simeq\pi/3$ and $\Delta\varphi_\circlearrowright\simeq4\pi/3$; for (011), $\Delta\varphi_\circlearrowleft\simeq\pi/2$ and $\Delta\varphi_\circlearrowright\simeq3\pi/2$.  
The origin of $\Delta\varphi$ will be discussed later. Nevertheless, such extraordinary findings suggest that $\vec{E}(t)$ can be well regarded as a broadband elliptically polarized THz pulse with its detailed characteristics depending on the pump polarization and the sample faces and, hence, has a defined chirality. According to the polarization trajectory ($S_{yz}(t)$,$S_x(t)$) in Fig.~\ref{fig:3D}\textbf{a} and \textbf{b}, chirality of the THz pulse can be instantaneously switched by varying the circular or linear polarization of pump light. 

\textbf{Polarization dependence of the THz signals.} To understand the peculiar THz wave emission from TaAs, it is necessary to elucidate the mechanism(s) generating the underlying time-resolved photocurrents. We measured the dependence of $S_x(t)$ and $S_{yz}(t)$ on the degree of circular polarization of the incident light, which can be controlled by rotating the quarter-wave plate by an angle $\theta$ (Fig.~\ref{fig:mainTHzresults}\textbf{a}). The experimental results were found to be well fitted by the following equation \cite{Osterhoudt_arXiv_2018,Ganichev_JPhys_2003,McIver_NNano_2011} 
\begin{eqnarray}
S_\lambda(t,\theta)&=&C_\lambda(t)sin2\theta+L_{1\lambda}(t)sin4\theta+L_{2\lambda}(t)cos4\theta \nonumber\\
& & +D_\lambda(t), 
\label{Eq:angle1}
\end{eqnarray}
where $\lambda$=$x$ or $yz$. $C_\lambda$ represents the contribution from helicity-dependent photocurrents. $L_{1\lambda}$ depends on the linear polarization and is phenomenologically associated with a quadratic nonlinear optical effect. $L_{2\lambda}$ and $D_\lambda$ arise from a thermal effect related to the light absorption. All four terms on the right side of Eq.~(\ref{Eq:angle1}) depend monotonically on the optical pump power, which agrees with our experimental observation (see Supplemental Material). 

\begin{figure*}
	\includegraphics[width=16cm]{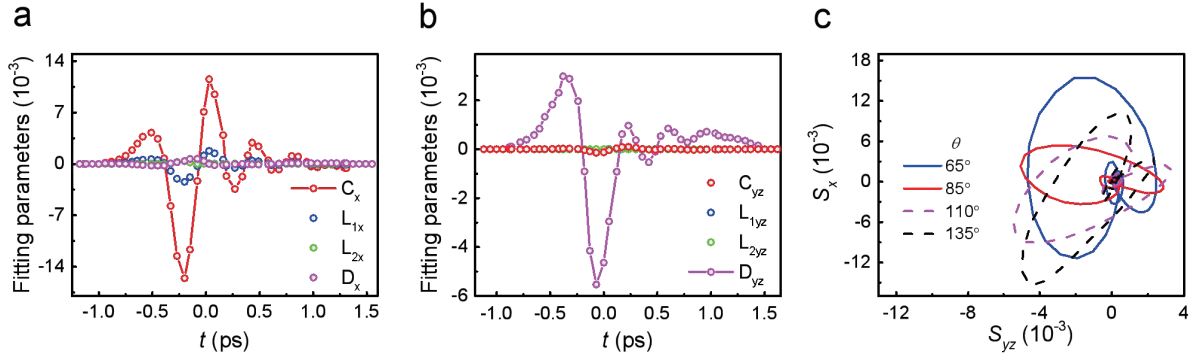}
	\caption{\label{fig:parameters} \textbf{a} and \textbf{b}. Time-dependent fitting parameters $C_\lambda$, $L_{1\lambda}$, $L_{2\lambda}$ and $D_\lambda$ ($\lambda=x,yz$) in Eq.~(\ref{Eq:angle1}). \textbf{c} shows the polarization trajectory ($S_{yz}(t)$,$S_x(t)$) under elliptically polarized pump light with different $\theta$. The solid and dashed curves represent opposite chiralities.} 
	%\vspace*{-0.4cm}
\end{figure*}

\begin{figure*}
	\includegraphics[width=16cm]{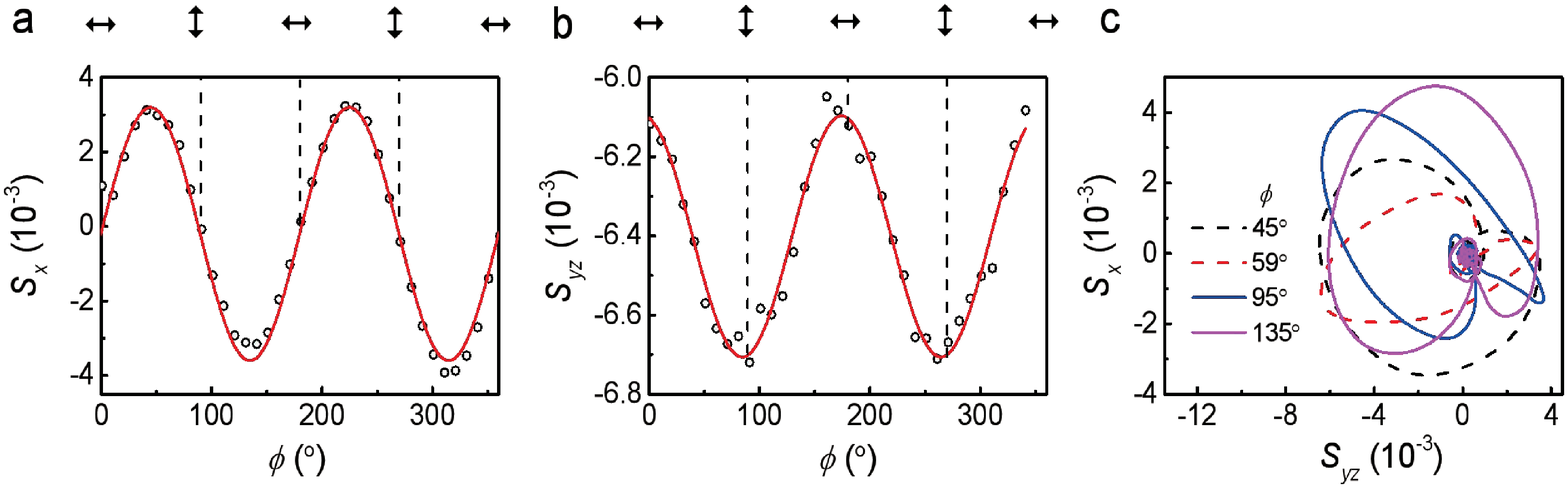}
	\caption{\label{fig:angle} \textbf{a} and \textbf{b} display the EO signals for $S_x(t=0.03$ ps) and $S_{yz}(t=-0.02$ ps) (near the peak values) as a function of the linear-polarization angle, $\phi$. The red solid lines show the fitted results. \textbf{c} shows the polarization trajectory ($S_{yz}(t)$,$S_x(t)$) for different linearly polarized pump light with several typical $\phi$. The solid and dashed curves represent opposite chiralities.} 
	%\vspace*{-0.4cm}
\end{figure*}

Figs.~\ref{fig:parameters}\textbf{a} and \textbf{b} display the time-dependent parameters $C_\lambda$, $L_{1\lambda}$, $L_{2\lambda}$ and $D_\lambda$, which were obtained by fitting the experimental $S(t,\theta)$ using the Eq.~\ref{Eq:angle1}. Fitting examples of $S_x(t)$ and $S_{yz}(t)$ as a function of $\theta$ can be found in the Supplemental Material. Based on our results, $S_x$ is unambiguously dominated by $C_x$, and has a non-negligible contribution from $L_{1x}$. Both the amplitude and phase of $S_x(t)$ change with $\theta$, while $L_{2x}$ and $D_{x}$ can be omitted. On the other hand, $S_{yz}(t)$ is dominated by a polarization-independent $D_{yz}(t)$. $C_{yz}$ plays a very small role, while $L_{1yz}$ and $L_{2yz}$ can be neglected. These results suggest that the ultrafast photocurrents leading to the THz signal $E_x$ (or $S_x$) is polarization-dependent (or $\theta$-dependent), in contrast to the polarization-independent thermally related photocurrent inducing $E_{yz}$ (or $S_{yz}$). Therefore, as demonstrated in Fig.~\ref{fig:angle}\textbf{c}, one can control the ellipticity and chirality of the elliptically polarized THz pulse by changing the quarter-wave plate angle $\theta$ (the elliptical polarization of the pump light). Realization of the broadband circularly polarized THz pulses also becomes possible, e.g., THz emission from the (011) face with $\Delta\varphi\simeq\pi/2$ (see Supplemental Material). 

Photocurrents arising from the linearly polarized light can be uncovered by measuring the dependence of $S_x$ and $S_{yz}$ on the linear polarization angle $\phi$, using a half-wave plate. The angle dependence of $S_x(t,\phi)$ near the peak values are shown in Figs.~\ref{fig:angle}\textbf{a} and \textbf{b}. We found that $S_x(\phi)$ can be well described by a second-order nonlinear optical process after considering the crystal symmetry, as demonstrated by the fitted curves in Figs.~\ref{fig:angle}\textbf{a} and \textbf{b} (see Appendix and Supplemental Material for detail fitting equations). On the other hand, $S_{yz}$ is only slightly modulated by the linear-polarization dependent signal and is dominated by a polarization-independent background. Similarly, we can manipulate the elliptically polarized THz pulse by changing the linear polarization state of the pump light, as illustrated in Fig.~\ref{fig:angle}\textbf{c}.  

\textbf{Ultrafast photocurrents in TaAs.} Observations of chiral broadband THz pulses indicate that the amplitude and phase of the ultrafast photocurrents can be fully controlled by polarized fs optical pulses. One can use the measured THz signals to quantitatively extract the ultrafast photocurrents (see Appendix), which are displayed in Figs.~\ref{fig:current}\textbf{a} and \textbf{b}. The data unambiguously demonstrates that switching of the current direction of $J_x(t)$ occurs instantaneously on a fs timescale using circularly or linearly polarized light, while $J_{yz}(t)$ is nearly unchanged for different polarized light. Along the time axis, $\vec{J}(t)$ shows spiral behavior. As a result, $\vec{J}(t)$ has chirality, which can be manipulated by the polarized pump light. Such results are rarely seen in conventional materials and, hence, lead to peculiar elliptically polarized ultrafast THz pulses. With regard to the dynamics of $J_x(t)$ and $J_{yz}(t)$, after an initial onset, $J_x(t)$ generally proceeds much faster than $J_{yz}(t)$. The former notably shows a strong oscillatory behaviour, which might be attributed to plasma oscillation (or plasmon) of the charge carriers. In fact, both previous FTIR \cite{Xu_PRB_2015} and our ultrafast optical transient reflectivity studies (see Supplemental Material) show that the Drude scattering time in TaAs has a timescale of $\sim$400 fs, which is consistent with the current relaxation time observed in Figs.~\ref{fig:current}\textbf{a} and \textbf{b}.

To determine the origin of $J_x$ and $J_{yz}$, we need to consider the mechanisms for the photocurrent generation. Microscopically, the ultrafast photocurrents can be generated during processes such as optical transitions, phonon- or impurity-scatterings, and electron-hole recombinations \cite{Braun_NC_2016}. Photocurrents induced by the optical transitions, occurring within the pulse duration, can in principle be controlled non-thermally in an ultrafast way \cite{Ganichev_JPhys_2003,McIver_NNano_2011,Chan_PRB_2017,Ma_NPhys_2017}. Of particular interest are the photocurrents due to the CPGE and LPGE \cite{Ganichev_JPhys_2003,Osterhoudt_arXiv_2018}, which are often respectively referred to as injection and shift currents \cite{Sipe_PRB_2000,Nastos_PRB_2006}. The former depends on the helicity of the pump light, while the latter is dependent on the crystal symmetry or the linear polarization state of the light. Based on the obtained sheet current densities, the injection currents due to CPGE play the main role in $J_x(t)$. We thus put the main focus on the injection currents.
    
\begin{figure*}
	\includegraphics[width=16cm]{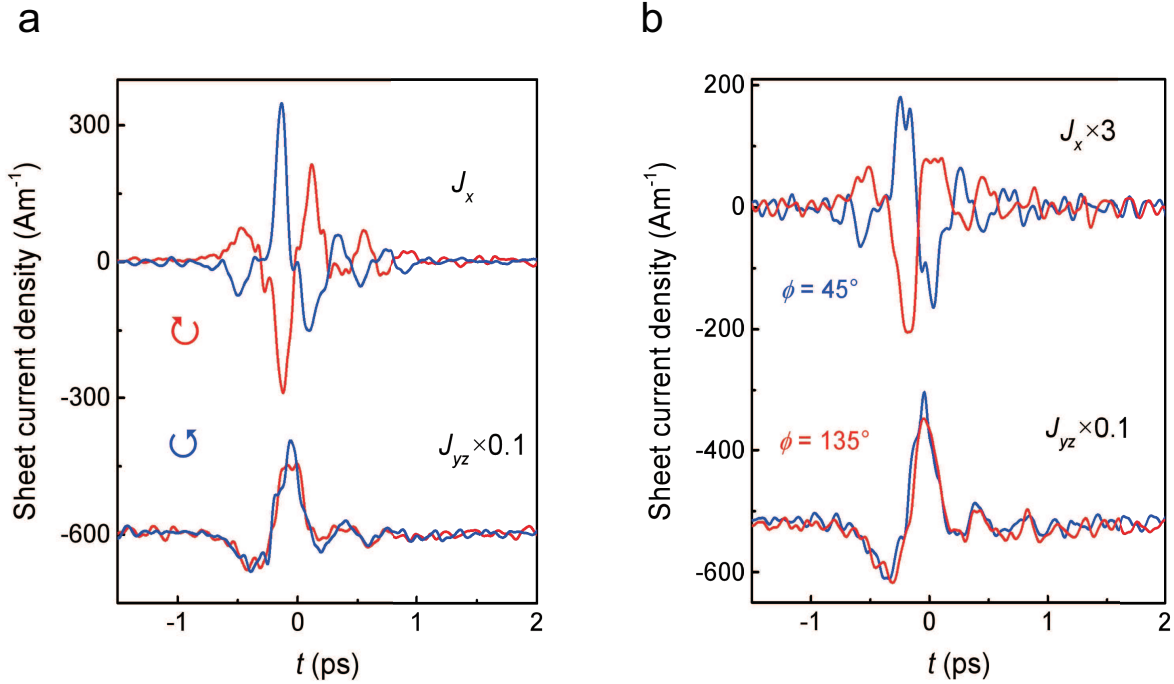}
	\caption{\label{fig:current} \textbf{a} and \textbf{b}. Extracted sheet photocurrent densities $J_x(t)$ and $J_{yz}(t)$ for different circular or linear pump polarization. Curves are offset for clarity.}
	%\vspace*{-0.4cm}
\end{figure*}

\begin{figure*}
	\includegraphics[width=16cm]{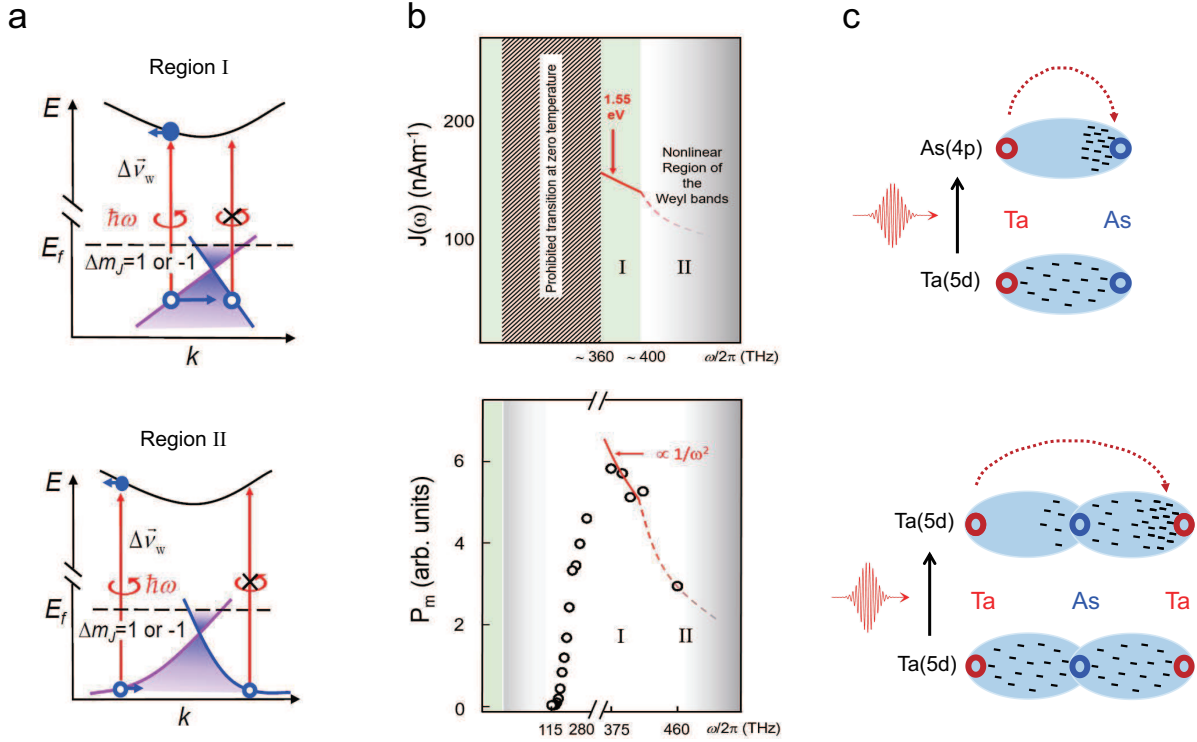}
	\caption{\label{fig:mechanisms} \textbf{a}. Schematic of CPGE. The asymmetric population of high-lying bands and depopulation of the Weyl cones leading to an average band velocity of $\Delta\vec{v}_w$ results in a nonzero charge current using circularly polarized light. During the optical transitions, the total angular momentum should be conserved, and its quantum number $m_J$ satisfies the relation: $\Delta m_J$=1 or -1, depending on the helicity of the pump pulse. The top is for the Weyl bands in the linear region (Region I). The bottom is for the Weyl bands in the nonlinear region (Region II), where the pump light has much higher photon energies. \textbf{b}. Theoretical and experimental results for the sheet current density and THz emission. The top displays our calculated sheet current density as a function of pump-light frequency (the red solid line). The calculated photocurrent is proportional to 1/$\omega$ (Region I). The bottom represents the measured THz peak power as a function of pump-light frequency (open dots). The red solid line in Region I is a fit using $1/\omega^2$ derived from our theory. The red dashed lines in Region II inside both figures are only for guideline. \textbf{c}. Schematic of the shift photocurrents (LPGE) is shown on the bottom. Due to the initial states being dominated by Ta 5$d$ orbitals and the final states contributed by both Ta 5$d$ and As 4$p$ orbitals \cite{Weng_PRX_2015}, the shift current is attributed to the ultrafast transfer of electron density along both the Ta-(As)-Ta and Ta-As bonds.}
	%\vspace*{-0.4cm}
\end{figure*}

\textbf{CPGE - Injection currents.} The scenario for CPGE is displayed in Fig.~\ref{fig:mechanisms}\textbf{a} \cite{McIver_NNano_2011}, where circularly polarized light introduces asymmetric population (depopulation) of the excited (initial) states complying with the angular momentum selection rules ($\Delta m_J=\pm1$). Due to band velocity differences among these states, an instantaneous charge current emerges, which is proportional to the average band velocity ($\Delta\vec{v}_w$). This scenario together with the cone tilting was employed to explain the helicity-dependent DC photocurrent in TaAs \cite{Chan_PRB_2017,Ma_NPhys_2017}. However, in contrast to those studies, where direct optical transitions only occurred within the Weyl cones due to usage of the long-wavelength infrared light with a photon energy of $\sim$120 meV, our experiments directly access the interband transitions between the Weyl cones and high-lying excited states above $E_f$ using excitation energies greater than $\sim$470 meV \cite{Weng_PRX_2015,Buckeridge_PRB_2016}. 

We thus carried out detailed theoretical calculations to clarify our observations. Note that the electromagnetic radiation is driven by the acceleration of the charge and is therefore proportional to the difference between the initial and final velocities of the charge resulting from its interaction with an external field. In our case, when the quasiparticle is excited by the laser light from a linear Weyl band with a large momentum-independent velocity $v = \partial E(q) / \partial q$ to a band with a much smaller velocity, the velocity difference is very large -- this makes Weyl semimetals ideal sources of induced radiation. The current then relaxes in the material over a typical time scale of $\sim 1$ ps, with this rapid deceleration of electric charges accompanied by electromagnetic radiation in the THz frequency range.
 
We assume the Hamiltonian in which a single Weyl cone makes a contribution is described by
\begin{equation}\label{hamilt}
\hat{H} =  \hbar v_{ia}\, \sigma_a\, q_i + \hbar v_{\text{t} i}\, \sigma_0\, q_i \equiv \hat{H}_{\text W} + \hat{H}_{\text t},
\end{equation}
where $\sigma_a$ are the Pauli matrices, $\sigma_0$ the identity matrix, $a$ is the pseudospin index and $i$ is the spatial index. The first term $\hat{H}_\text{W}$ contains information about the chirality and the velocity of the Weyl fermion, and the second term $\hat{H}_\text{t}$ describes  the tilt in the direction determined by the constant vector $\vec v_{\text{t}}$; $q_i$ is the quasiparticle's momentum measured from the position of the Weyl node; $v_{ia}$ is the velocity matrix. The interaction with the electromagnetic field $\vec{A}$ is obtained from (\ref{hamilt}) through the Peierls substitution $\vec{q} \to \vec{q} - \frac{e}{\hbar}\vec{A}$; this leads to the electromagnetic interaction Hamiltonian $\hat{H}_{\text EM}$
\begin{equation}\label{hamilt_em}
\hat{H}_\text{EM} =  -e v_{ia}\, \sigma_a\, A_i - e\, v_{\text{t} i}\, \sigma_0\, A_i \equiv \hat{H}_{\text W EM} + \hat{H}_{\text t EM}. 
\end{equation}
Here, the second term (which we denote by $\hat{H}_{\text t EM}$) is diagonal in spin space and does not contribute if one considers transitions between the Weyl bands; however, it will in general contribute once other nonlinear bands are excited.

Using the Hamiltonian (\ref{hamilt_em}), the induced electric current density can now be readily computed basing on two physical assumptions: i) we can neglect the band velocity of an excited band compared to the velocity on the Weyl band \cite{Weng_PRX_2015,Buckeridge_PRB_2016}, so the energy of the excited band can be assumed to be approximately independent of momentum; and ii) once the photons enter the material, they will induce an excitation with unit probability. Assumption ii) may not be realistic due to other excitations induced by the photons, e.g., the shift photocurrents discussed in later sections. Based on these assumptions and using Fermi's golden rule, we can write the current density integrated over the penetration depth (the DC sheet current density) as
\begin{align}
\vec{J}(\omega,\vec{k}_p, \vec{\varepsilon}) &= \int \vec{j}(\omega,\vec{k}_p, \vec{\varepsilon})\, \du z \nonumber \\
&= \frac{-eI}{\hbar \omega}\,  \frac{\sum_{l} \tau_l a_l \int \frac{\du^3q}{2\pi^3}\, \delta(E_{l-}(q)-E_{l0})\, (0-\vec{v}_{l-}(q))\, \sum_i \lvert \braket{s_{li}| \hat{H}_{\text EM}|q_{l-}}\rvert^2}{\sum_{l} a_l \int \frac{\du^3q}{2\pi^3}\, \delta(E_{l-}(q)-E_{l0})\, \sum_i \lvert \braket{s_{li}| \hat{H}_{\text EM}|q_{l-}}\rvert^2}\label{jdef},
%\label{eq:Currentcalculation}
\end{align}
where the summation over $l$ is the summation over the 24 Weyl cones of TaAs. $\omega, \vec{k}_p, \vec{\varepsilon}$ and $I$ are the frequency, momentum, polarization and intensity of the pump light entering the material. $v_-(q)$ is the band velocity; $E_-$ is the energy on the valence Weyl band; $E_{0}$ is the energy of the excited band minus $\hbar\omega$; $\ket{s_i}$ is the spin state of the excited band, and $\tau_l$ is the current relaxation time. The relaxation time appears in Eq.\eqref{jdef} because Fermi's golden rule yields the number of transitions per unit time and we have to integrate it over the lifetime of the current. The prefactor of $\frac{I}{\hbar\omega}$ is the flux of photons entering the material. The factor $a_l$ is the spatial overlap between the wave functions of the Weyl band and the excited band. We assume that the difference between $E_{-}(q)$ and the Fermi energy is much greater than the temperature. We neglect the momentum transfer from light to the quasiparticle due to the small incident angle $\Theta$.

After performing the integrals, Eq.\eqref{jdef} will be of the form 
\begin{align}
J_i(\omega,\vec{k}_p, \vec{\varepsilon})  
&= \frac{-eI}{\hbar \omega}\,  \frac{\sum_{l} \tau_l a_l \chi_l N^i_{(l)j} L^j}{\sum_{l} a_l D^{ij}_{(l)} \varepsilon_i \varepsilon^*_j},
\end{align}
where $N^i_{(l)j}$ and $D^{ij}_{(l)}$ are tensors that depend on the dispersion relations of the cones, and are independent of the frequency of light as long as the Weyl bands are linear; $\chi_l = \pm 1$ is the chirality of each Weyl cone (the $+$ and $-$ signs correspond to right- and left-handed cones, respectively); $\vec{L} = \iu\vec{\varepsilon}\times\vec{\varepsilon}^*$ is the angular momentum per photon. For circularly polarized light, $\vec{L}=\pm\hbar\hat{k}_p$.

TaAs has tetragonal symmetry, i.e. 4-fold rotational symmetry about an axis and reflection symmetry about 4 planes containing that axis. It also has time reversal symmetry. This means the 24 Weyl points exist as a set of 8 ($W_1$) and a set of 16 ($W_2$), with the cones in each set related by the crystal symmetries. Chirality is invariant under rotations and time reversal, and flips sign under reflections. This means each set has an equal number of left and right handed cones. If we take the sum over a set of cones, the symmetric components of $N^i_{(l)j}$ cancel and the only non-canceling contribution is from $N^x_{(l)y} - N^y_{(l)x}$. Therefore, the chiral photocurrent in Eq.\eqref{jdef} is $\vec{J}\propto \pm\hat{k}_p \times \hat{c}$, where $+$ and $-$ signs refer to the right- and left-handed polarizations of light. 

The chiral photocurrent is perpendicular to both the [001] crystallographic axis and the momentum of light $\vec{k}_p$, and it reverses sign for different circular polarization or opposite direction of $\hat{c}$-axis (see also the Supplemental Material). The chiral nature of Weyl cones cannot contribute to the LPGE and hence produce a current along the [001] axis. We consider the light incident approximately normal to the (112) face of the crystal. In this case, the chiral photocurrent is along the $[\bar{1}10]$ direction.

For excitation light with a wavelength of 800 nm, the numerical evaluation of Eq.\eqref{jdef} (see Supplemental Material for more details) yields a value of $J_1 = + 940$ nA/m for the contribution of the 8 Weyl cones $W_1$ to the sheet current density. For the classification of the Weyl cones with different chiralities in TaAs, we follow the supplementary materials of Ref.~\cite{Ma_NPhys_2017} and use an optical penetration depth of 25 nm (see Supplemental Material). For the 16 Weyl cones $W_2$, the sheet current density $J_2 = - 1340$ nA/m. Due to the lack of detailed information about the probabilities of excitation for the two sets of cones $W_1$ and $W_2$, we are only able to reliably obtain the range of the sheet current density from $-1340$ nA/m to $940$ nA/m. If we assume that 8 Weyl cones $W_1$ and 16 Weyl cones $W_2$ are excited with equal probabilities, we obtain -580 nA/m (for right circular polarization) or +580 nA/m (for left circular polarization). Experimentally, the peak value of the helicity-dependent $J_x(t)$ reaches almost 350 A/m. Considering the photocurrent flows over a time $<\tau_l>\simeq$400 fs at a repetition rate of $f_{rep}$=1 KHz, we evaluated an equivalent DC sheet current density via $\overline{J_x}\simeq$max($J_x$)$f_{rep}<\tau_l>$, which gives $\overline{J_x}\sim$140 nA/m. This value is consistent with the our theoretical result, although there exists some discrepancy between their numbers, which could partly be due to i) our rough estimation of the experimental $\overline{J_x}$ and ii) our assumption that the Weyl bands absorb all of the photons and contribute solely to the helicity-dependent photocurrents. In fact, observation of the shift photocurrents indicates that this approximation is not very accurate. Extraordinarily, we found that the sheet current density is inversely proportional to the frequency $\omega$ of the incident light. This dependence holds for frequencies above the threshold ($\sim$360 THz) for excitation of the high lying non-Weyl bands and below the energy ($\sim$400 THz) at which non-linearity of the Weyl bands sets in, as indicated by the Regions I and II in Fig.~\ref{fig:mechanisms}\textbf{a}. 

According to the above description, the peak power of THz radiation for incident light with the photon energy $\hbar\omega$, defined by $P_m\propto|E_{max}|^2$, is also proportional to the square of DC current density  $\vec{J}(\omega,\vec{k}_p, \vec{\varepsilon})$ and inversely proportional to the duration of the current:
\begin{equation}
P_m\propto \frac{1}{\tau^2} \lvert J \rvert^2\propto \frac{1}{\omega^2}.
\end{equation} 
Because the sheet current density is inversely proportional to the frequency $\omega$ of the incident light, the radiation power is proportional to $1/\omega^2$ within the frequency range described above. As shown in the Region I of Fig.~\ref{fig:mechanisms}\textbf{b}, our experimental data consistent well with the calculations. We note, however, that the CPGE-induced THz signal or photocurrent in experiments does not suddenly drop to zero in the frequency range, where our theory predicts that the optical transitions are prohibited at zero temperature (the shaded region), and instead is manifested by gradual decrease. Such behavior may arise from the band tailing effect due to the non-zero temperature and defects in the sample \cite{Stern_PR_1966,Mieghem_RMP_1992}.     

Therefore, we demonstrate both experimentally and theoretically that for excitation light with high photon energies Weyl physics is the key determinant for the observed giant helicity-dependent photocurrents and the resultant coherent THz emission. This is because the excitation from the Weyl band to the high-lying band is accompanied by a large and rapid change in the effective velocity of the charged quasiparticles. The microscopic mechanism is quite different from previous findings in the WSMs upon light excitations with small photon energies \cite{Chan_PRB_2017,Zhang_PRB_2018,Ma_NPhys_2017}, where the Weyl band tilting plays the pivotal role.

One would argue that helicity-dependent photocurrents may arise from other mechanisms, such as the circular photon drag effect (CPDE) \cite{Shalygin_PRB_2016} or the spin-galvanic effect (SGE) \cite{Ganichev_JPhys_2003}. The CPDE current requires additional transfer of the light momentum along the charge current direction. This effect will be irrelevant for the case here with a small incidence angle. For the SGE current, its decay is determined by the spin relaxation time, $\tau_s$ \cite{Ganichev_JPhys_2003}. Based on our time-resolved Kerr rotation measurement, the observed $J_x(t)$ cannot be explained by SGE since $\tau_s\simeq$60 fs is too small (see Supplemental Material). 

\textbf{LPGE - Shift currents.}  LPGE, on the other hand, depends on the crystal symmetry or the linear polarization of light. Its induced charge current is also called the shift current \cite{Ganichev_JPhys_2003,Osterhoudt_arXiv_2018,Sipe_PRB_2000}, which occurs when the electron density distribution of the excited state is spatially shifted with respect to the initial states during the optical transitions (see Fig.~\ref{fig:mechanisms}\textbf{c}). LPGE is absent for systems with inversion symmetry under a quadratic nonlinear optical process \cite{Ganichev_JPhys_2003} and is thus also referred to as the bulk photogalvanic effect\cite{Osterhoudt_arXiv_2018}. Our experimental data are well explained by the phenomenological Eq.~(\ref{eq:LPGE}) derived from this effect. Here, because the Weyl cones dominated by the Ta $5d$ orbitals and the high-lying bands include both Ta $5d$ and As $4p$ orbitals \cite{Weng_PRX_2015}, the associated ultrafast electron transfer should occur along both Ta-(As)-Ta and Ta-As bonds (see Fig.~\ref{fig:mechanisms}\textbf{b}). Clearly, the crystal symmetry is embedded in the anisotropy of these bonds. By contrast, for low photon energies, e.g., $\sim$120 meV, the electron transfer directly between Ta and Ta along the Ta-(As)-Ta bond dominates the shift currents. The colossal shift photocurrents found in Ref.~\cite{Osterhoudt_arXiv_2018} should belong to this case. Based on our results, the related sheet current density has a peak value of $\sim$68 A/m, which corresponds to an equivalent DC sheet current density of $\sim$28 nA/m. Considering the power density of pump light, this value is consistent with the findings in Ref.~\cite{Osterhoudt_arXiv_2018}, where a distinct photon energy of $\sim$120 meV was used. The observed large shift current response might indicate giant non-linear third-rank tensors $\xi_{\lambda\mu\nu}$ present in Eq.~(\ref{eq:LPGE}), similar to previous second harmonic generation studies in this material \cite{Wu_NPhys_2017}. Our experiments thus demonstrate that the ultrafast shift current response is also significant, and provides an additional control degree of freedom for ultrafast THz pulses using linearly polarized light on the fs timescale. 

\textbf{Photo-thermal currents.} $J_{yz}$ is largely independent of any pump polarization. Therefore, discussing its mechanism can basically exclude the CPGE and LPGE since they only show a little contribution to the signal. Based on Eq.~(\ref{Eq:angle1}), the dominance of $D_{yz}$ in the $S_{yz}$ (or $E_{yz}$) signal indicates $J_{yz}$ has a thermal origin. Its relatively slow response in the time domain further suggests that scattering processes following the initial optical transitions are involved during the generation of $J_{yz}$. Possible candidates are the photo-Dember effect \cite{Dekorsy_PRB_1996} and carrier drift, both of which depend strongly on the phonon or impurity scatterings. An estimation of $J_y(t)$ and $J_z(t)$ can be obtained using the data by varying the incident angle $\Theta$ (see Supplemental Material). Their magnitudes are close to that of the shift current.    

Therefore, the distinct polarization dependent $J_x(t)$ and $J_{yz}(t)$ arise from different physical mechanisms, non-thermal and thermal, respectively. A phase difference between these two components is expected. As a result, the chiral ultrafast photocurrent $\vec{J}(t)$ emerges, as evidenced by our experiment. Such phase difference directly determines the observed $\Delta\varphi$ between $E_x(t)$ and $E_{yz}(t)$. An estimation of $\Delta\varphi$ can be roughly made by $\Delta\varphi\simeq\bar{\Omega}<\tau_{ep}>$, where $\bar{\Omega}$ and $\tau_{ep}$ are the angular THz frequency and the electron-phonon scattering time, respectively. Using the frequency ($\sim$1.8 THz) at the peak magnitude of the Fourier transform and the average $<\tau_{ep}>$ value of $\sim$400 fs, we obtain $\Delta\varphi\simeq1.4\pi$, which is close to the values measured values, i.e. $4\pi/3$ for (112) and $3\pi/2$ for (011). Potential anisotropy of the detail electron-phonon scatterings corresponding to different faces may cause such various phases. 

\section{Conclusions} 
The demonstrated ultrafast generation and control of chiral charge currents in TaAs offer unique opportunities for novel THz emission and THz spintronics. The mechanism of generating controllable elliptically/circularly polarized broadband THz pulses using WSMs is fundamentally different from previous methods. The intrinsic optical chiralities, i.e. the optical/chirality selection rules in WSM, which do not apply for previous THz emitters, such as ZnTe and LiNbO$_3$. The simplicity of polarization control is extremely powerful and useful for a wide range of applications. Other advantages include the low cost in sample preparation and the high THz emission efficiency. We further believe that our observation will benefit the study of other novel phenomena led by the Weyl physics, such as the quantized CPGE \cite{Juan_NC_2017}, and the Weyl-orbit effect\cite{Zhang_NC_2017}.\newline

\textbf{Acknowledgments} This research is supported by the National Natural Science Foundation of China (Grant No. 11734006 and No. 11674369), the Frontier Science Project of Dongguan (No. 2019622101004), the National Key Research and Development Program of China (No. 2016YFA0300600 and No. 2018YFA0305700), the Science Challenge Project of China (TZ2016004), the National Postdoctoral Program for Innovative Talents (Projects No. BX201700012, funded by China Postdoctoral Science Foundation), and the CAS interdisciplinary innovation team. This work is also supported in part by the U.S. Department of Energy under Awards DE-SC-0017662 (E.J.P. and D.E.K.), DE-FG-88ER40388 (S.K. and D.E.K.), and DE-AC02-98CH10886 (D.E.K.).\newline

\textbf{Author contributions} Y.G. built the terahertz emission setup and performed the measurements with the help of Y.L., Y.S. and X.C. Z.L. prepared the TaAs sample. Y.G., Y.Q., and J.Q. analyzed the experimental data. S.K., E.J.P. and D.E.K. performed theoretical computation of the helicity-dependent photocurrents. M.L. and H.W. provided theoretical assistance. Y.G., M.L. and J.Q. wrote the manuscript with contributions from all the co-authors. J.Q. conceived the experiments and supervised the project.\newline  

\textbf{Competing financial interests} The authors declare that they have no competing financial interests.\newline
    
\section{Methods}
\subsection{Sample details}
Large-size high-quality TaAs crystals with regular shapes and shiny facets were grown by the chemical vapor transport (CVT) method. High-purity elemental tantalum, arsenic, and iodine with a molar ratio of 1:1:0.05 were filled into a silica ampule under argon. The ampule was evacuated to a pressure below 1 Pa, sealed quickly by flame to avoid the loss of iodine and arsenic, and then heated gradually from room temperature to 1000 $^\circ$C to get TaAs polycrystalline. Afterwards the ampule was put to a temperature gradient from 1020 $^\circ$C to 980 $^\circ$C where the CVT proceeded for 2 weeks and single crystals were obtained. More details can be seen elsewhere \cite{zhilin_CGD_2016}. The crystal structure and orientations were determined by x-ray diffraction method and the average stoichiometry was confirmed by energy-dispersive x-ray spectroscopy. TaAs crystallizes in a body-centered tetragonal unit cell where Ta and As atoms are six coordinated to each other. The corresponding lattice constants are $a=b=$3.43 \textup{\AA} and $c$=11.64 \textup{\AA}, and the space group is I41md (No. 109). 

\subsection{Experimental setup}
In our experiments, the sample was excited by the laser pulses from a Ti:sapphire amplifier (repetition rate 1 KHz, duration $\sim$80 fs, and centre wavelength 800 nm) under $\sim3^\circ$ or 45$^\circ$ angle of incidence. Other wavelengths of the excitation light come from an optical parametric amplifier, which produces similar pulse duration (78$\pm$3 fs). The beam diameter is $\sim$1.5 mm (full-width at half intensity maximum). A typical pump power of 25 mW was used. The THz electric field was detected by electro-optic sampling, with probe pulses from the same laser co-propagating with the terahertz field through an electro-optic crystal, which is the ZnTe(110) with a thickness of 0.4 mm. Measurements were performed at room temperature in a dry-air environment with relative humidity$<5\%$. All data were collected in the linear regime, i.e. amplitude of THz field increases linearly with the pump power (see Supplemental Material).

A quarter-wave ($\lambda/4$) or half-wave ($\lambda/2$) plate mounted in a computer-controlled rotation stage is employed to tune the polarization state of the optical pulses just before they reach to the sample. A THz wire-grid polarizer (field extinction ratio of 10$^{-2}$) allows us to measure the $s$ and $p$ components $E_x$ and $E_{yz}$ of the THz electric field separately, thereby disentangling current components $J_x$ and $J_{yz}$. The latter is a linear combination of $J_y$ and $J_{z}$ \cite{Shan_AP_2004}.    

\subsection{Extraction of the THz electric fields}
To extract the emitted THz electric field $\vec{E}(t)$ directly above the sample surface from the measured electro-optic THz signal $\vec{S}(t)$, there is a linear relationship between these two quantities \cite{Kampfrath_NNano_2013}. For instance, in the frequency domain, the THz field component $E_x$ and the corresponding signal $S_x$ are connected by the total transfer function $h(\Omega)$ through the simple multiplication\cite{Kampfrath_NNano_2013}
\begin{equation}
S_x(\Omega)=h(\Omega)E_x(\Omega),
\end{equation}
where $h(\Omega)=h_{det}(\Omega)h_{prop}(\Omega)$ includes the detector response $h_{det}(\Omega)$ and the transfer function $h_{prop}(\Omega)$ of the THz wave from the sample to the detection crystal. The same relationship is valid for $E_{yz}$ and $S_{yz}$. Details of the transfer functions are shown in the Supplemental Material.

\subsection{Extraction of the ultrafast photocurrents}
In order to obtain the source current $\vec{J}(t)$ from the THz electric field $\vec{E}(t)$ measured directly above the sample surface, we make use of the following generalized Ohm’s law \cite{Braun_NC_2016,Shan_AP_2004}:
\begin{eqnarray}
E_x(\Omega)=-\frac{Z_0}{cos\Theta+\sqrt{n^2-sin^2\Theta}}J_x(\Omega)\\
E_{yz}(\Omega)=-\frac{Z_0sin\Theta}{n^2cos\Theta+\sqrt{n^2-sin^2\Theta}}J_{yz}(\Omega).
\end{eqnarray}
Here, $\Omega$ is the THz frequency, $Z_0(\simeq$377 Ohm) is the vacuum impedance, $n$ is the refractive index of TaAs at THz frequency (see Supplemental Material), $\Theta$ is the angle of incidence, and
\begin{equation}
J_{yz}=J_z(\Omega)-\frac{\sqrt{n^2-sin^2\Theta}}{sin\Theta}J_y(\Omega)
\end{equation} 
is a weighted sum of the currents flowing along the $\hat{y}$ and $\hat{z}$ directions. The inverse Fourier transformation of the resulting current spectra yields the currents in the time domain.

\subsection{Photocurrents due to the LPGE (shift currents)}
Phenomenologically, the non-local transient photocurrent $\vec{j}(\vec{r},\Omega)$ at THz frequency $\Omega$ due to the LPGE can be described by a quadratic nonlinear optical process \cite{Ganichev_JPhys_2003,Ivchenko_SPS_2008,Kampfrath_NNano_2013,Braun_NC_2016,Mills_book_2012} 
\begin{eqnarray}
j_\lambda(\vec{r},\Omega)=2\sum_{\mu\nu}\int_{\omega>0} d\omega\xi_{\lambda\mu\nu}(\vec{r};\omega+\Omega,\omega)F_{\mu}f^*_{\nu},
\label{eq:LPGE}
\end{eqnarray}
where $\lambda$, $\mu$ and $\nu$ stand for the Cartesian coordinates $\hat{x}$, $\hat{y}$ and $\hat{z}$. $\xi_{\lambda\mu\nu}$ is the third-rank pseudo-tensor. $\vec{F}$ and $\vec{f}$ are the complex-valued pump-field Fourier amplitudes at frequencies $\omega+\Omega$ and $\omega$ originating from the fs optical pump pulse. Due to $\omega>>\Omega$, $|\vec{F}(\omega+\Omega)|\simeq|\vec{f}(\omega)|$.

For TaAs with inversion symmetry broken, there are three independent nonvanishing elements of $\xi_{\lambda\mu\nu}$ \cite{Wu_NPhys_2017}: $\xi_{zzz}$ , $\xi_{zxx}=\xi_{zyy}$ and $\xi_{xzx}=\xi_{yzy}=\xi_{xxz}=\xi_{yyz}$. They are defined in the coordinates for (001) face, where $\hat{x}$, $\hat{y}$ and $\hat{z}$ are parallel to the unit cell axises $\hat{a}$, $\hat{b}$, and $\hat{c}$, respectively. For the coordinates of other faces, i.e. (011) and (112), they will follow the transformation of the rotation matrix. Therefore, $j_\lambda(\vec{r},\Omega)$ due to the LPGE will be different at different faces (see Supplemental Material).

\end{document}